\documentclass{jetpl}
\twocolumn

\lat

\title{New limits on di-nucleons decay into invisible channels}

\rtitle{New limits on di-nucleons decay\ldots}

\sodtitle{New limits on di-nucleons decay into invisible channels}

\author{V.\,I.\,Tretyak\/\thanks{e-mail: tretyak@kinr.kiev.ua},
V.\,Yu.\,Denisov,
Yu.\,G.\,Zdesenko}

\rauthor{V.\,I.\,Tretyak, V.\,Yu.\,Denisov, Yu.\,G.\,Zdesenko}

\sodauthor{Tretyak, Denisov, Zdesenko}

\address{Institute for Nuclear Research, MSP 03680 Kiev, Ukraine}

\dates{12 December 2003}{*}

\abstract{Data of the radiochemical experiment [E.L.~Fireman, 1978] with 1.7 t of
KC$_2$H$_3$O$_2$, accumulated deep underground during $\simeq$1 yr, were
reanalyzed to set limits on di-nucleons ($nn$ and $np$) decays into
invisible channels (disappearance, decay into neutrinos, etc.). The obtained
lifetime bounds
$\tau_{np}>2.1\times 10^{25}$ yr and
$\tau_{nn}>4.2\times 10^{25}$ yr (at 90\% C.L.)
are better (or competitive) than those
established in the recent experiments.}


\begin{document}

\maketitle

More than three decades searches for the proton decay, which is predicted by
the Grand Unified Theories, continue to be one of the most important and
intriguing subjects in quest of the effects beyond the Standard Model of
elementary particles \cite{Lan81}. Up to now, only lifetime limits were
established for such processes, being on the level of $\tau >10^{30}-10^{33}$
yr for the nucleons decay into particles, which can strongly or
electromagnetically interact with the nuclei contained in the detector's
sensitive volume \cite{PDG02}. Recently, the interest increased to nucleon
decays into so called ``invisible'' channels (which are complementary to
conventional ones \cite{PDG02}) when nucleon or pair of nucleons decay into
some weakly interacting particles (for example, neutrinos) or disappear. The
last possibility is related with theories describing our world as
four-dimensional brane embedded into higher-dimensional structure \cite
{Ynd91,Ark98,Rub01}. According to \cite{Rub01}, disappearance of particles
into extra dimensions is a generic property of matter. Searches for
disappeared energy and/or momentum in particles' collision are planned with
accelerators at high energies \cite{Hew02}. An experiment to search for
disappearance of orthopositronium is discussed in \cite{Gni03}. Perspectives
to search for invisible decays of neutrons and di-neutrons in $^{12}$C with
the 1000 t KamLAND detector are examined in \cite{Kam03}, and sensitivities
of future 1000 t lead perchlorate detector for $n$ disappearance in $^{35}$%
Cl and $^{208}$Pb are considered in \cite{Boy03}.

As for the to-date status, the most stringent limits for nucleons and
di-nucleons decay into invisible channels have been known from the experiments
performed during few last years (all bounds are given with 90\% C.L.):

(1) $\tau _p>3.$5$\times $10$^{28}$ yr -- from number of free neutrons which
could be created in result of $p$ disappearance in deuterium nuclei ($d=pn$%
), which are contained in 1000 t of D$_2$O of the SNO apparatus \cite{Zde03};

(2) $\tau _p>3$.9$\times $1$0^{29}$ yr and $\tau _n>3$.9$\times $1$0^{29}$
yr -- from number of $\gamma $ quanta with $E_\gamma =6-7$ MeV which will be
emitted in deexcitation of $^{15}$O or $^{15}$N after $n$ or $p$
disappearance in $^{16}$O nucleus in 1000 t of the SNO heavy water \cite
{Ahm03};

(3) $\tau _{pp}>5$.0$\times $1$0^{25}$ yr and $\tau _{nn}>4.$9$\times $1$%
0^{25}$ yr -- from the search for decay of radioactive nuclei ($^{10}$C, $%
^{11}$Be and $^{14}$O) created after $pp$ and $nn$ disappearance in $^{12}$%
C, $^{13}$C and $^{16}$O nuclei in liquid scintillator (4.2 t of C$_{16}$H$%
_{18}$) and water shield (1000 t) of the BOREXINO Counting Test Facility
\cite{Bac03};

(4) $\tau _{np}>3.$2$\times $1$0^{23}$ yr -- from the search for decay of $%
^{134}$I created in result of $np$ disappearance in $^{136}$Xe \cite{Ber03}.

In order to improve the $\tau _{np}$ limit, we reanalyze here the data of
the old radiochemical experiment \cite{Fir78} where the daughter nuclide $%
^{37}$Ar was searched for as a possible product of the $p$ or $n$
disappearance in $^{39}$K. The target, 1710 kg of potassium acetate KC$_2$H$%
_3$O$_2$ which contains 9.7$\times $10$^{27}$ atoms of $^{39}$K, was exposed
deep underground (the Homestake mine, 4400 m w.e.) during more than 1 year.
The production rate of $^{37}$Ar, extracted from the target and detected due
to its radioactive decay $^{37}$Ar $\to $ $^{37}$Cl ($T_{1/2}=35$ d), for
the last 3.5 months period was measured as $0.$3$\pm $0$.6$ atom/day. On
this basis authors have accepted the limit on production rate of $^{37}$Ar
as one atom/day and have calculated the restrictions on the $p$ and $n$
lifetimes \cite{Fir78,Ste78}. For example, after the $p$ decay in $_{19}^{39}
$K, the nucleus $_{18}^{38}$Ar will be created, as a rule being in an
excited state (unless the disappeared $p$ was on the outermost shell). The
authors estimated that in 22.2\% of cases additional neutron will be emitted
from $_{18}^{38}$Ar in deexcitation process giving rise to $_{18}^{37}$Ar
nucleus \cite{Fir78,Ste78}. Similarly, after the $n$ disappearance in
initial $_{19}^{39}$K, produced $_{19}^{38}$K emits $p$ with 20.4\%
probability, which will also result in the $_{18}^{37}$Ar nucleus. From
these values, accounting for 19 protons and 20 neutrons in the $_{19}^{39}$%
K, the limits $\tau _p=\tau _n=1.$1$\times $1$0^{26}$ yr were set \cite
{Fir78,Ste78}.

However, the same data can be used to calculate the $\tau _{np}$ limit, just
noticing that simultaneous disappearance of the $np$ pair in $_{19}^{39}$K
also will produce the $_{18}^{37}$Ar nucleus. The corresponding limit on
lifetime can be derived by using the formula:
\begin{equation}
\lim \tau =N_{nucl}\times N_{obj}^{\mathrm{eff}}\times t/\lim S,
\end{equation}
\noindent where $N_{nucl}$ is the number of $^{39}$K nuclei; $N_{obj}^{%
\mathrm{eff}}$ is the ``effective'' number of objects (here $np$ pairs)
whose disappearance in the parent nucleus will result in creation of the
daughter nuclide; $t$ is the time of measurements; and $\lim S$ is the
number of effect's events which can be excluded at a given confidence level
on the basis of the experimental data.

According to the Feldman-Cousins procedure \cite{PDG02,Fel98}, the measured
value of $^{37}$K production rate $S/t=0.$3$\pm $0.$6$ atom/day results in
the limit $\lim S/t=$ 1.28 atom/day at 90\% C.L. Conservatively supposing
only \textit{one} $np$ pair (for 1 unpaired proton in the $_{19}^{39}$K
nucleus; disappearance of the outermost proton and neutron on nucleons shell
in parent nucleus will produce daughter in a non-excited state) and using
the eq. (1) with $N_{nucl}=9.$7$\times $1$0^{27}$, we obtain the following $%
np$ lifetime limit:

\begin{center}
$\lim \tau _{np}=2.$1$\times $1$0^{25}$ yr at 90\% C.L.
\end{center}

In addition, the $\tau _{nn}$ bound can be also determined: disappearance of
the $nn$ pair from $_{19}^{39}$K nucleus will give $_{19}^{37}$K which
quickly decays again to $_{18}^{37}$Ar with $T_{1/2}=1.2$ s \cite{ToI96}%
\footnote{%
Unfortunately, disappearance of the $pp$ pair results in creation of stable
nucleus $^{37}_{17}$Cl, and, thus, cannot be investigated in this approach.}%
. The number of objects, $N_{obj}^{\mathrm{eff}}$, can be calculated in the
following way \cite{Bac03,Eva77,Ber00}. After disappearance of neutrons with
binding energies $E_{n1}^b(A,Z)$ and $E_{n2}^b(A,Z)$ in $(A,Z)$ nucleus, the
excitation energy of the $(A-2,Z)$ daughter, $E_{exc}$, can be approximated
as $E_{exc}=E_{n1}^b(A,Z)+E_{n2}^b(A,Z)-2S_n(A,Z)$, where $S_n(A,Z)$ is the
binding energy of the least bound neutron in the $(A,Z)$ nucleus. In the
process of deexcitation of the $(A-2,Z)$ daughter only $\gamma $ quanta can
be emitted when the value of $E_{exc}$ is lower than the binding energy of
the least bound nucleon in the $(A-2,Z)$ nucleus: $E_{exc}<S_N(A-2,Z)$,
where $S_N(A-2,Z)=\min \{S_n(A-2,Z),~S_p(A-2,Z)\}$\footnote{%
Higher excitations of daughter nucleus will result in deexcitation process
with emission of mostly $n$, $p$, etc., instead of $\gamma$ quanta, and give
not the $(A-2,Z)$ nucleus but isotopes with lower $A$ and $Z$ values.}.
Under this condition we receive the restriction on the values of the
neutrons binding energies: $E_{n1}^b(A,Z)+E_{n2}^b(A,Z)<2S_n(A,Z)+S_N(A-2,Z)$%
.

Values of the separation energies $S_n$ and $S_p$ were taken from \cite
{Aud95}. Single-particle energies $E_n^b(A,Z)$ for neutrons in the $%
_{19}^{39}$K nucleus were calculated with the WSBETA code \cite{Cwi87} using
the Blomqvist-Wahlborn parameterization of the Woods-Saxon potential \cite
{Blo60}. Calculated value of the neutron separation energy $S_n^{\mathrm{calc%
}}=13.08$ MeV is in good agreement with the experimental value $S_n^{\mathrm{%
exp}}=13.07$ MeV \cite{Aud95}. We conservatively suppose that contributions
to the effective number of objects, $N_{obj}^{\mathrm{eff}}$, give only
paired neutrons (i.e. neutrons with equal values of all quantum numbers,
except for the magnetic quantum number), and neglect contributions from
other neutrons. Taking into account that the binding energies of such
particles are equal, the appropriate equation is as follows: $%
2E_n^b(A,Z)<2S_n(A,Z)+S_N(A-2,Z)$. This condition gives only 2 $nn$ pairs
whose disappearance from $_{19}^{39}$K will produce relatively low-excited
daughter $_{19}^{37}$K, which emit only $\gamma $ quanta (hence, can not be
transformed to nucleus with $A<37$ in result of ejection of additional
nucleons). Substituting the values $N_{nucl}=9.$7$\times $1$0^{27}$, $%
N_{obj}^{\mathrm{eff}}=2$ and $\lim S/t=1.28$ 1/day in eq. (1) one gets

\begin{center}
$\lim \tau _{nn}=4.$2$\times $1$0^{25}$ yr at 90\% C.L.
\end{center}

In conclusion, reanalysis of the data of the radiochemical experiment of
Fireman \cite{Fir78} allows us to establish the limits: $\tau _{nn}>4$.2$%
\times $1$0^{25}$ yr and $\tau _{np}>2.$1$\times $1$0^{25}$ yr at 90\% C.L.
The $\tau _{nn}$ value is near the same as that given recently by the
BOREXINO Collaboration ($\tau _{nn}>4.$9$\times $1$0^{25}$ yr \cite{Bac03}),
while the obtained value for $\tau _{np}$ is two orders of magnitude higher
than that set in \cite{Ber03} and is the most restrictive up-to-date limit
for the $np$ decays into invisible channels.

\end{document}